%% file: main.tex
\documentclass{osa-article}

\usepackage{gensymb}
\usepackage{xcolor}
\usepackage{multirow}
\usepackage{subfiles}
\journal{oe}
\articletype{Research Article}
\usepackage{chapterbib}
\usepackage{chemformula}

\newcommand{\mts}[1]{_{\scriptscriptstyle\mathit{#1}}}
\begin{document}

\input{Filters1.tex}

\bibliography{Master-2}
\end{document}

%% file: Filters1.tex
\title{Ultra-narrowband polarization insensitive transmission filter using a coupled dielectric metal metasurface}
\author{Arvind Nagarajan\authormark{1,2,*}, Koen van Erve\authormark{3} and Giampiero Gerini\authormark{1,2}}

\address
{\authormark{1}Electromagnetics Group, Technische Universiteit Eindhoven (TU/e), 5600 MB Eindhoven, The Netherlands\\
\authormark{2}Optics Department, Netherlands Organization for Applied Scientific Research (TNO), Stieltjesweg 1, 2628 CK Delft, The Netherlands\\
\authormark{3}Department of Applied Physics, Technische Universiteit Eindhoven (TU/e), 5600 MB Eindhoven, The Netherlands\\ }

\email{\authormark{*}a.nagarajan@tue.nl} 

% ==============================================================================
\begin{abstract*}
    
A coupled dielectric-metal metasurface (CDMM) consisting of amorphous silicon (a-\ch{Si}) rings and subwavelength holes in \ch{Au} layer separated by a \ch{SiO2} layer is presented. The design parameters of the CDMM is numerically optimized to have a polarization independent peak transmittance of 0.55 at 1540 nm with a Full Width at Half Maximum ($FWHM$) of 10 nm. The filter also has a 100 nm quite zone with $\sim 10^{-2}$ transmittance. A radiating two-oscillator model reveals the fundamental resonances in the filter which interfere to produce the electromagnetically induced transparency (EIT) like effect. Multipole expansion of the currents in the structure validates the fundamental resonances predicted by the two-oscillator model. The presented CDMM filter is robust to artifacts in device fabrication and has performances comparable to a conventional Fabry-P\'{e}rot filter. However, it is easier to be integrated in image sensors as the transmittance peak can be tuned by only changing the periodicity resulting in a planar structure with a fixed height.
\end{abstract*}

\section{Introduction}
Ultra-narrowband spectral filtering are crucial for a variety of applications like LiDAR \cite{Schwarz2010}, optical satellite communication \cite{Sodnik2010}, gas detection \cite{Gibson2017} and multispectral imaging \cite{Lapray2014}. Traditionally, ultra-narrowband filters are based on Distributed Bragg Reflectors (DBR) constituting a Fabry-P\'{e}rot cavity \cite{Weis1988}. The height of the cavity determines the peak wavelength, while the number of alternating dielectric layers in the DBR determines the bandwidth, amplitude and the Full Width at Half-Maximum ($FWHM$) . A typically DBR filter with a $FWHM$ of 10 nm consists of $\sim$10 layers \cite{Macleod2017}. A recent alternative to the DBR filters is the Volume Bragg Hologram (VBH) filters \cite{Lumeau2010, Lumeau2006} produced by holographic recording in a photosensitive substrate. However, VBH filters are only suitable for low power applications and the photosensitive substrate bleaches with time.\par
Spectral filtering is also resonantly possible in metasurfaces by interference of two different excitation pathways leading to Electromagnetically Induced Transparency (EIT) \cite{Fleischhauer2005}. EIT has been observed in both plasmonic resonators \cite{Zhang2008,Tassin2009,Vafapour2017,Rana2018} and all-dielectric resonators \cite{Yang2014,Zhang2014}. Plasmonic EIT filters \cite{Ebbesen1998,Ortuno2011,Shah2018} typically consists of subwavelength holes in a single metal layer. However, at optical frequencies plasmonic filters have low transmission efficiency due to losses and have a wide bandwidth limiting the quality factor ($Q$, defined as the ratio between resonance frequency (f) and its $FWHM: \dfrac{f}{\Delta f}$) to around 10 \cite{Shah2018}. \par
All-dielectric resonators, however, have low radiative losses and low absorption loss enabling high $Q$ resonances. Dielectrics with high permittivity such as \ch{Ge} or \ch{Si} can generate electric dipoles, magnetic dipoles and higher order modes due to Mie resonances \cite{Zhao2009}. Among the various multipoles generated in a dielectric, toroidal dipole resonance is of interest as it produces an ultra-high $Q$-factor ($>10^3$) \cite{Algorri2019,Tasolamprou2016,Fan2013,Kaelberer2010,Savinov2014,Basharin2015,Tuz2018,Gupta2016,Han2018}. Dielectric resonators with toroidal resonances cannot be used as-is for spectral filtering as the quite zone around the sharp resonance is typically $<\lambda/30$. The ultra-high $Q$ factor of the toroidal resonance is also extremely sensitive to the geometry. A small change ($\sim$1\%) in the geometry (due to artefacts in device fabrication) can bring down the $Q$-factor by several orders of magnitude. Unfortunately, research on toroidal resonances so far have only focused on increasing the $Q$-factor, without paying much attention to increasing the quite zone.\par
In this work, we envision a coupled dielectric-metal metasurface (CDMM) filter inspired by recent works on stacked metasurfaces \cite{Wan2015,Taubert2012,Zhu2015,Menzel2016,Han2019} which present additional degrees of freedom in achieving high-$Q$ resonances. The CDMM filter, consists of amorphous silicon (a-\ch{Si}) rings \cite{vandeHaar2016} coupled in the near-field to a plasmonic layer consisting of subwavelength holes in \ch{Au} with a \ch{SiO2} spacer layer. The CDMM filter combines the advantages of a high-$Q$ resonance offered by the a-\ch{Si} rings with the excellent out-of-band performance of the plasmonic layer. The designed filter has a $Q$-factor of 154 at 1540 nm, with a 100 nm ($\lambda/15$) quite zone. A radiating two-oscillator model reveals the fundamental resonances which interfere to produce EIT. Multipole decomposition of the currents in the CDMM filter further confirms the dipole-dipole interaction predicted by the two-oscillator model.
\section{Numerical Simulations}
We envision a coupled dielectric-metal metasurfaces (CDMM) filter shown in \autoref{fig:Filters_schematic} consisting of a-\ch{Si} rings and subwavelength holes in \ch{Au} layer separated by a \ch{SiO2} layer. Numerical simulations were carried out using finite element method (FEM) with a commercially available software COMSOL Multiphysics\textsuperscript{\textregistered}. The Johnson and Christy database \cite{Johnson1972} was used for \ch{Au} permittivity, Malitson model \cite{Malitson1965b} was used for \ch{SiO2} permittivity, and Pierce and Spicer database was used for a-\ch{Si} permittivity \cite{Pierce1972}. In the 3D FEM simulation, Floquet boundary conditions are employed in both $x$ and $y$ directions, and ports (with sufficient diffraction orders) are employed in the $z$ direction. A plane wave is normally incident with electric-field along $y$ direction as shown in \autoref{fig:Filters_schematic}. 
\input{Figures1/fig_Filters_schematic}
\par
The key design parameters of the CDMM filter mentioned in \autoref{fig:Filters_schematic} were optimized to have peak transmittance at 1540 nm. \ch{Si} ring has an outer diameter ($D_1$) of 1110 nm, width ($W$) of 220 nm and height ($H$) of 250 nm. The \ch{SiO2} spacer has a thickness ($H_s$) of 750 nm. The ground \ch{Au} layer has a thickness ($H_g$) of 200 nm, with an air hole of diameter ($D_2$) 740 nm. The unit cell is placed in a square lattice with a periodicity ($\Lambda$) of 1500 nm.\par

The simulated normal incidence transmission (including diffraction) is plotted in a log scale in \autoref{fig:Transmittance_Comsol} as it readily gives the optical density ($OD$). Transmittance ($T$) is directly related to OD as $T = 10^{-OD}$.  The peak transmittance of 0.55 with a full width at half maximum ($FWHM$) of 10 nm is reported at 1540 nm ($Q$ factor 154.19). The designed CDMM filter has a quite zone in the 1500--1600 nm spectral range. The $Q$ factor can be increased at the cost of both the peak transmittance and the quite zone bandwidth. Diffraction orders are present for wavelengths below 1500 nm, and the effective transmittance including all diffraction orders is represented by the red curve in \autoref{fig:Transmittance_Comsol}. A broad peak with a transmittance of 0.35 is observed at 1623 nm.  The 100 nm bandwidth quite zone has $OD_1$ bandwidth (i.e. $T < 0.1$)  of 31 nm and 53 nm to the left and the right of the peak respectively, and $OD_2$ bandwidth (i.e. $T < 0.01$)  of 24 nm and 21 nm at the extreme left and right ends respectively. The quite zone performance of the designed metamaterial filter is hence comparable to that of a DBR filters. The peak intensity unfortunately is not 1 due to resonant losses in the \ch{Au} layer. The absorbance of the DMHM filter is shown in \autoref{fig:Absorbance_comsol}. A peak absorbance of 0.44 is observed at 1540nm. 
\input{Figures1/fig_Transmittance_Comsol}
\input{Figures1/fig_Absorbance_comsol}
\section{Mechanism}
The peak transmittance of the CDMM filter at 1540 nm is due to a  hybrid mode of the a-\ch{Si} ring and the \ch{Au} layer leading to an EIT effect. The moduli of electric field for the resonance wavelength of 1540 nm along various cut planes of the CDMM filter are shown in \autoref{fig:E_Field_1540nm}. The white arrows indicate the field components. There is a coupling between the a-\ch{Si} rings and the \ch{Au} layer as revealed by the electric field lines along the center of the CDMM filter in the $xz$ plane shown in \autoref{fig:E_Field_1540nm}(a). The four symmetric lobes of the electric field around the a-\ch{Si} ring seen in \autoref{fig:E_Field_1540nm}(b) suggests that the resonance is lattice assisted. The electric field lines along the air gaps both in a-\ch{Si} rings and in \ch{Au} layer resembles an electric dipole.\par
A radiative two-oscillator model \cite{Tassin2012,Hu2017} is used to decouple the hybrid mode of the CDMM filter into fundamental resonances which interfere to produce an EIT like effect. The modes of both resonators are assumed to be bright as they can both be directly excited by the incoming light. The position of an electron in the a-\ch{Si} resonator is described by $ x_{\ch{Si}}(t) $ with a resonance frequency of $ \omega_{\ch{Si}} $ and a damping factor of $ \gamma_{\ch{Si}} $. Similarly, the electron position in the \ch{Au} resonator is described by $ x_{\ch{Au}}(t) $ with a resonance frequency of $ \omega_{\ch{Au}} $ and a damping factor $ \gamma_{\ch{Au}} $.
\input{Figures1/fig_E_Field_1540nm}
Considering an external excitation $ f_{1}(t) $ , the coupling between the electrons are given by the following equations: (see appendix B for elaboration): 
\input{Equations1/eq_EquationOfMotion}
Here, $\Omega$ is the complex coupling coefficient. Assuming a time harmonic excitation force $ f(t)=\tilde{f}e^{-i\omega t} $, the equations of motion can be solved assuming the solutions to have the form $ x_{\ch{Si}}(t)=X_{\ch{Si}}e^{-i\omega t} $ and  $ x_{\ch{Au}}(t)=X_{\ch{Au}}e^{-i\omega t} $
\input{Equations1/eq_EquationOfMotionSolution}
\par
As the thickness of the resonators are deep subwavelength, the electric current density ($J$) in the resonators can be homogenized as a sheet of effective surface conductivity ($ \sigma_{e} $)  \cite{Tretyakov2015}. i.e. $ J=-in_{s}\omega \left( X_{\ch{Si}} + X_{\ch{Au}} \right) =\sigma_{e}E_{s}  $, where $n_{s}$ and $ E_{s} $ denote the average electron density and spatially averaged electric field on the current sheet respectively.\par
Assuming $ \tilde{f} \propto E_{s} $ surface conductivity is given by the following equation:  
\input{Equations1/eq_SurfaceConductivity}
Finally, the transmission coefficient is related to the surface conductivity by following equation \cite{Li2018}:
\input{Equations1/eq_TransmissionCoefficient}
\input{Figures1/fig_Transmission_fitting}
\par
The numerically obtained transmittance is fit with $ \bigl\lvert T\bigr\rvert^{2} $ obtained by \autoref{eq:TransmissionCoefficient} in \autoref{fig:Transmission_fitting}. The fundamental resonances $ \omega_{\ch{Si}} $ and $ \omega_{\ch{Au}} $ were restricted to 1140--1300 THz (i.e. 1450--1650 nm wavelength). The damping terms $ \gamma_{\ch{Si}} $ and $ \gamma_{\ch{Au}} $ was bounded to positive values (negative values suggest gain medium). The average electron density $n_{s}$ was also bounded to positive values. The retrieved fit parameters are shown in the inset. a-\ch{Si} rings have a fundamental resonance at 1500 nm, and the \ch{Au} layer has a fundamental resonance at 1629 nm. It is interesting to observe that the damping of a-\ch{Si} resonator is 10 MHz, while that of the \ch{Au} layer is 4.187 THz. The fit shows an excellent agreement with the numerical simulations. The CDMM filter can hence be simplified as a system of two coupled dipoles with fundamental resonances at 1500 nm and 1650 nm respectively. \par
The currents in the CDMM filters are decomposed into various cartesian multipole moments to validate the two-oscillator model in the 1450--1650 nm spectral range. The polarization induced current density $\mathbf{J}(r)$ of a nanoparticle in a host medium is related to the electric field $\mathbf{E}(r)$ by the expression: \input{Equations1/eq_inlineCurrentDensity} in cartesian basis assuming $e^{-i\omega t}$ convention for the time harmonic electromagnetic fields. \par
The electric ($p$) and  magnetic ($m$) dipolar moments of the nanoparticle can be expressed as  \cite{Gurvitz2019}: 
\input{Equations1/eq_ElectricDipole}
\input{Equations1/eq_MagneticDipole}
The toriodal dipole moments can be split into Toridal electric dipole ($T^{(e)}$) and Toroidal magnetic dipole ($T^{(m)}$). 
\input{Equations1/eq_ToroidalEDipole}
\input{Equations1/eq_ToroidalMDipole}
The electric (${\bar{\bar{Q}}}^{(e)}$) and magnetic (${\bar{\bar{Q}}}^{(m)}$) quadrupoles are symmetrized and traceless tensors expressed as : 
\input{Equations1/eq_ElectricQuadrapole}
\input{Equations1/eq_MagneticQuadrapole}
The toroidal electric (${\bar{\bar{T}}}^{(Qe)}$) and magnetic (${\bar{\bar{T}}}^{(Qm)}$) quadrupoles are also symmetrized and traceless tensors expressed as: 
\input{Equations1/eq_ToroidalEQuadrapole}
\input{Equations1/eq_ToroidalMQuadrapole}
%Here, $j,k$ represents the coordinates $\lbrace 1,2,3\rbrace$. $\delta_{jk}$ is the Kronecker delta function. The scattered power can also be decomposed into various multipoles by the following equations. As the scattered power by $T^{(m)}, {\bar{\bar{T}}}^{(Qe)} \text{ and } {\bar{\bar{T}}}^{(Qm)}$ are quite low, they are included as a correction factor in $m, {\bar{\bar{Q}}}^{(e)} \text{ and } {\bar{\bar{Q}}}^{(m)}$ respectively. Octupoles, and other higher order modes are ignored.
\begin{list}{}{}
    \setlength{\itemsep}{0pt}%
    \setlength{\parskip}{0pt}%
    \item $j,k$ represents the coordinates $\lbrace 1,2,3\rbrace$ for the coordinates $\lbrace x, y,z\rbrace$
    \item $\delta_{jk}$ is the Kronecker delta function
    \item $\vec{\mathbf{r}}$ is the position vector, given by $\vec{\mathbf{r}}=r_x\hat{x}+r_y\hat{y}+r_z\hat{z}$
    \item $r^2$ is the modulus square of the position vector, given by $r^2=r_{x}^{2}+r_{y}^{2}+r_{z}^{2}$
\end{list}
The scattered power can also be decomposed into various multipoles by the following equations. As the scattered power by $T^{(m)}, {\bar{\bar{T}}}^{(Qe)} \text{ and } {\bar{\bar{T}}}^{(Qm)}$ are quite low, they are included as a correction factor in $m, {\bar{\bar{Q}}}^{(e)} \text{ and } {\bar{\bar{Q}}}^{(m)}$ respectively. Octupoles, and other higher order modes are ignored.
\input{Equations1/eq_PScatEDipole}
\input{Equations1/eq_PScatTEDipole}
\input{Equations1/eq_PScatMDipole}
\input{Equations1/eq_PScatEQpole}
\input{Equations1/eq_PScatMQpole}

The scattering cross-section $\sigma\mts{scat}$ is then obtained by normalizing the scattered power $P\mts{scat}$ to the incident energy flux.
\input{Equations1/eq_scatteringSection}
Here, $E\mts{inc}$ is the incident electric-field. Finally, the scattering efficiency ($Q\mts{scat}$) is obtained by normalizing the scattering cross-section with the geometric cross-section.
\input{Equations1/eq_ScatteringEfficiency}
\input{Figures1/fig_Filters_multipole_expansion}
\par
The polarization induced current density $\mathbf{J}(r)$ of the a-\ch{Si} ring and the \ch{Au} layer were numerically evaluated. The scattering efficiency was then computed with origins at the center of the a-\ch{Si} ring and at the center of the \ch{Au} layer as shown in \autoref{fig:Filters_multipole_expansion}. As inferred, a-\ch{Si} has a sharp scattering peak at 1500 nm, which is predominantly a magnetic quadrupole moment. A sharp toroidal electric dipole peak is also observed at 1500 nm. Au layer has a narrow scattering peak at 1540 nm and a broad scattering peak at 1629 nm, both of which are predominantly dominated by magnetic quadrupoles.\par
A weak electric dipole is also observed at 1629 nm. Although magnetic quadrupoles dominate the scattering in both a-\ch{Si} rings and in \ch{Au} layer, the peak transmittance at 1540 nm is due to dipole-dipole interactions as predicted by the two-oscillator model. The two quadrupoles do not interact with each other as the quadrupole---quadrupole interactions are extremely local and decays quickly compared to a dipole---dipole interaction, with the interaction energies inversely proportional to the 5$^{\text{th}}$ power of distance between them \cite{Knipp1938}. \par
The moduli of electric fields for the wavelengths 1500 nm and 1629 nm along various cut planes are shown in \autoref{fig:E_Field_1500nm_1629nm}. The white arrows indicate the field components. At 1500 nm, there is a strong electric field concentration at  the air gap inside the a-\ch{Si} ring as seen in \autoref{fig:E_Field_1500nm_1629nm}(a, b) due to toroidal dipole caused by circulating magnetic field in the a-\ch{Si} ring as indicated by the dotted black curve in \autoref{fig:E_Field_1500nm_1629nm}(c). At 1629 nm, there is a strong electric field concentration at  the air gap inside the \ch{Au} layer as seen in \autoref{fig:E_Field_1500nm_1629nm}(e, f). It is to be noted here that the field concentration at the air gap is maximum at this wavelength. There is also some field concentration at the a-\ch{Si} ring at this wavelength. The peak transmittance at 1540 nm is a hybridization between the toroidal dipole at the air gap inside a-\ch{Si} ring at 1500 nm, and the effective electric dipole at the air gap inside the \ch{Au} layer at 1629 nm validating the two-oscillator model.
\input{Figures1/fig_E_Field_1500nm_1629nm}
\par
The dependence of the transmittance of the CDMM filter on the axial offsets between the a-\ch{Si} ring and the \ch{Au} layer was numerically simulated and reported in \autoref{fig:Transmission_offset}. The offset along the polarization axis has little influence on the transmittance, the peak shifts by 1 nm for an offset of 100 nm as shown in \autoref{fig:Transmission_offset}(a). However, offset perpendicular to polarization axis has a drastic effect on the transmittance. The transmittance has two diverging peaks for offsets greater than 10 nm as seen in \autoref{fig:Transmission_offset}(b). This behavior can be easily understood by considering the CDMM filter as a system of two coupled dipoles. Offset along the polarization axis does not affect the coupling between the two, while offset perpendicular to the polarization axis weakens the coupling. This also confirms that the CDMM filter is not a system of two coupled quadrupoles as the quadrupole-quadrupole interactions should have an identical response for both $x$ and $y$ axial offsets. 
\input{Figures1/fig_Transmission_offset}
\par
The robustness of the CDMM filter is numerically simulated by varying the crucial design parameters $D_1$ and $D_2$ as reported in \autoref{fig:Transmission_Robustness}. Although toroidal resonances are extremely sensitive to the geometry, we observe a marginal change in the transmittance when the a-\ch{Si} ring diameter was changed by $\pm$30 nm as reported in \autoref{fig:Transmission_Robustness}(a). The transmittance has negligible dependence on the diameter of the subwavelength holes in the \ch{Au} layer as inferred from \autoref{fig:Transmission_Robustness}(b). Clearly, the coupling between the sharp toroidal resonance with the broad dipole resonance of \ch{Au} layer makes the CDMM filter more robust to artefacts in device fabrication.\par
\input{Figures1/fig_Transmission_Robustness}
The circular symmetry of the geometry ensures that the CDMM filter is not dependent on the polarization, i.e. the azimuth angle ($ \phi $).  However, the vertical separation between the a-\ch{Si} ring and the \ch{Au} layer creates a launch angle ($\theta$) dependence. \autoref{fig:Angle_sweep} shows the numerically simulated launch angle dependence on the transmittance of the CDMM filter. The transmittance has two diverging peaks for a launch angle greater than 0.5\degree  . The CDMM filter is hence extremely sensitive to the launch angle.\par
\input{Figures1/fig_Angle_sweep}
The peak transmittance of the CDMM filter can be tuned just by changing the periodicity ($ \Lambda $) of the unit-cell as the resonances are strongly dependent on the periodicity.  \autoref{fig:Period_Sweep} plots the transmittance as a function of unit cell periodicity. As seen, the CDMM filter can be easily tuned in the C-band infrared. The quite zone around the central peak has a bandwidth of $\sim$100 nm. 
\input{Figures1/fig_Period_Sweep}

\section{Conclusions}
A novel coupled dielectric-metal metasurfaces (CDMM) polarization insensitive transmission filter design is presented in this work. The filter consists of arrays of a-\ch{Si} rings and air holes in \ch{Au} layer separated by a \ch{SiO2} spacer layer.  The designed filter has a peak transmittance of 0.55 at 1540 nm with a $FWHM$ of 10 nm. The ultra-narrowband transmission peak is surrounded by a quite zone of 100 nm with a $OD_2$ bandwidth of 45 nm at the extreme ends and a $OD_1$ bandwidth of 84 nm around the central peak. The ultra-narrowband transmission is due to the coupling between the sharp toroidal resonance in a-\ch{Si} ring with the broad dipolar resonance of the subwavelength air holes in \ch{Au} layer. The coupling between the metasurfaces are validated using a radiative two-oscillator model and modal decomposition showing excellent agreement with numerical simulations. The CDMM filter is robust to artefacts in device fabrication. The CDMM filter is resistant to offsets between the two layers along the polarization axis, and is stable till 10 nm offset perpendicular to the polarization axis. The CDMM filter is extremely sensitive to the launch angle due to the vertical separation between the two layers. The central peak of the filter can be easily tuned in the entire C-band infrared spectral range by changing the periodicity keeping all other design parameters fixed. Multi-spectral filtering can hence by attained in a single device with a fixed height by combining multiple unit cells with varying periodicity. The proposed CDMM filter is hence easier to be integrated into a CMOS image sensor.  

\section*{Acknowledgments}
This work was funded by TU/e through the MELISSA PhD project. The authors would like to thank Jonas Berzinš for comments on the manuscript. 

%% file: Figures1/fig_Filters_schematic.tex
\begin{figure}[h!]
 \centering
  \includegraphics[width=0.7\textwidth]{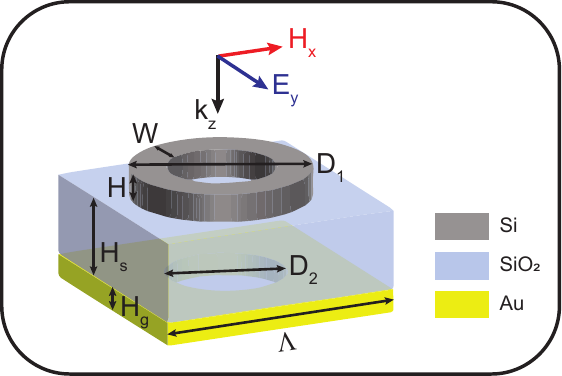}
\caption{Schematic of the proposed CDMM filter highlighting the key design parameters.}
\label{fig:Filters_schematic}
\end{figure}

%% file: Figures1/fig_Transmittance_Comsol.tex
\begin{figure}[h!]
 \centering
  \includegraphics[width=0.7\textwidth]{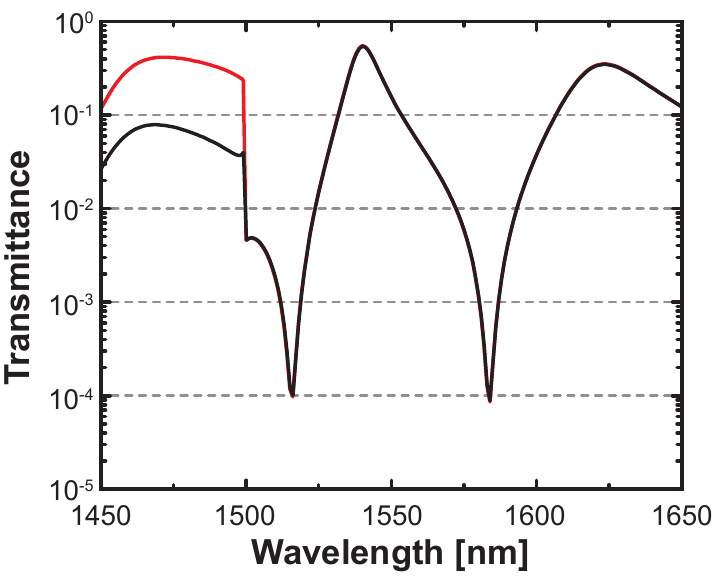}
\caption{Numerically simulated normal incidence transmittance of the designed CDMM filter under planewave illumination. The black curve represents $0^{\text{th}}$ order transmittance, while the red curve represents total transmittance including all diffraction orders present.}
\label{fig:Transmittance_Comsol}
\end{figure}

%% file: Figures1/fig_Absorbance_comsol.tex
\begin{figure}[h!]
 \centering
  \includegraphics[width=0.7\textwidth]{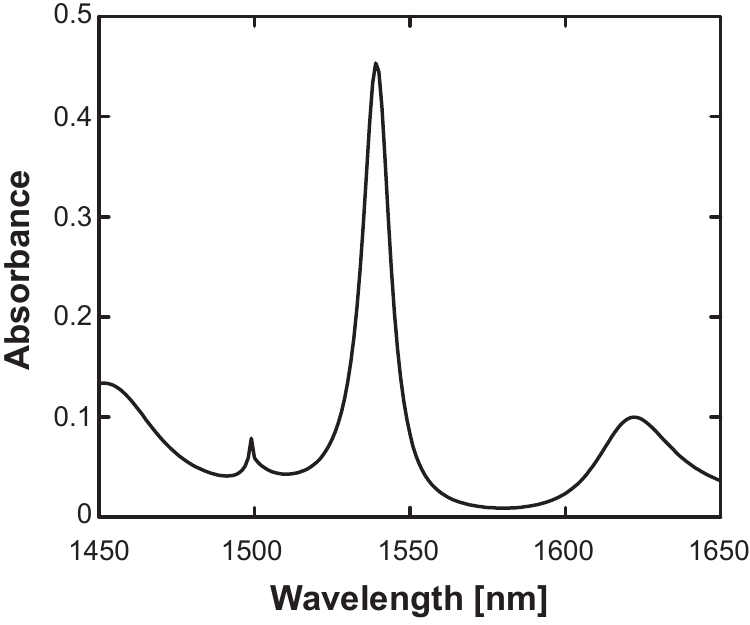}
\caption{Numerically simulated normal incidence absorbance of the designed CDMM filter for a normal incident plane wave illumination.}
\label{fig:Absorbance_comsol}
\end{figure}

%% file: Figures1/fig_E_Field_1540nm.tex
\begin{figure}[h!]
 \centering
  \includegraphics[width=1\textwidth]{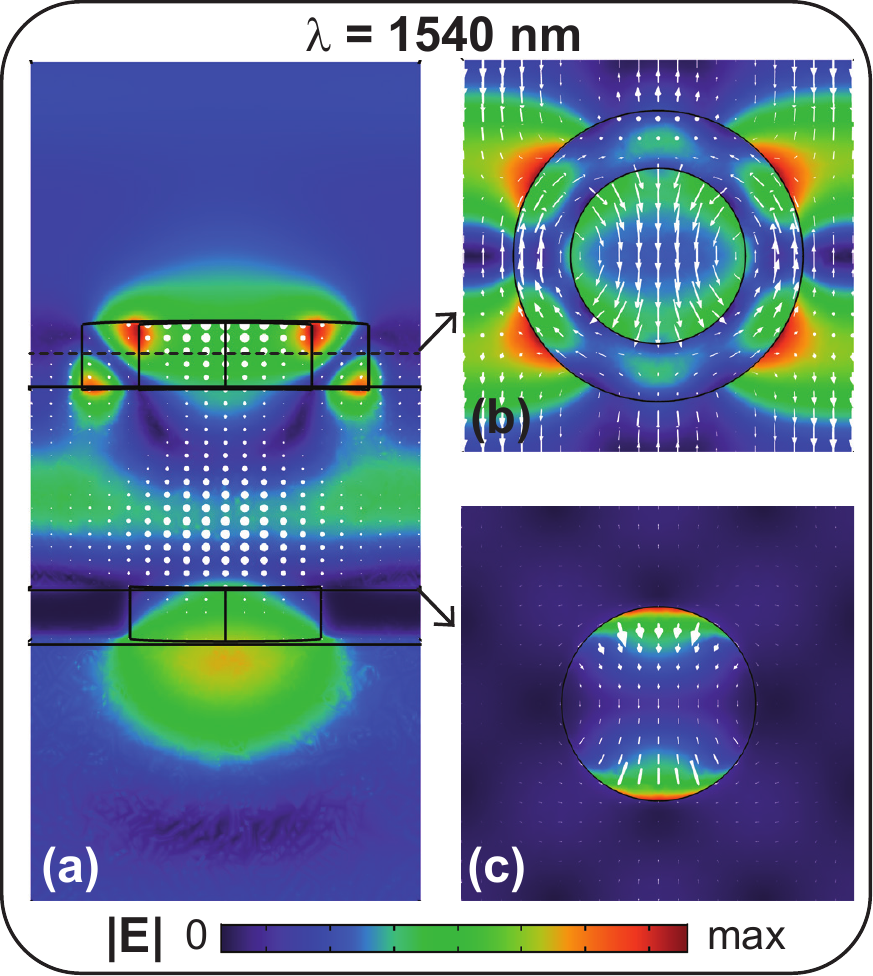}

\caption{Moduli of electric fields of the CDMM filter under a $y$-polarized planewave illumination of 1540 nm wavelength. The white arrows, with lengths proportional to the magnitude, indicate the field components. (a)
vertical cut in the center of the CDMM filter along $xz$ plane. Horizontal cuts along the (b) center of a-\ch{Si} ring and (c) top of \ch{Au} layer.}
\label{fig:E_Field_1540nm}
\end{figure}

%% file: Equations1/eq_EquationOfMotion.tex
\begin{subequations}
\label{eq:EquationOfMotion}
\begin{align}
        \ddot{x}_{\ch{Si}}(t)+\gamma_{\ch{Si}}\dot{x}_{\ch{Si}}(t)+\omega_{\ch{Si}}^2x_{\ch{Si}}(t)-\Omega^2e^{i\varphi}x_{\ch{Au}}(t)&=f_{1}(t)\label{eq:EquationOfMotionSi} \\
        \ddot{x}_{\ch{Au}}(t)+\gamma_{\ch{Au}}\dot{x}_{\ch{Au}}(t)+\omega_{\ch{Au}}^2x_{\ch{Au}}(t)-\Omega^2e^{i\varphi}x_{\ch{Si}}(t)&=f_{1}(t)  \label{eq:EquationOfMotionAu}
\end{align}
\end{subequations}

%% file: Equations1/eq_EquationOfMotionSolution.tex
\begin{subequations}
\label{eq:EquationOfMotionSolution}
\begin{align}
	X_{\ch{Si}}&= \dfrac{\Omega^2+\left( \omega^{2}_{\ch{Au}} -\omega^{2} -i\omega\gamma_{\ch{Au}} \right) }{\left( \omega^{2}_{\ch{Si}} -\omega^{2} -i\omega\gamma_{\ch{Si}} \right)\left( \omega^{2}_{\ch{Au}} -\omega^{2} -i\omega\gamma_{\ch{Au}} \right)-\Omega^{4}}\tilde{f} \label{eq:EquationOfMotionSolutionSi} \\
	X_{\ch{Au}}&= \dfrac{\Omega^2+\left( \omega^{2}_{\ch{Si}} -\omega^{2} -i\omega\gamma_{\ch{Si}} \right) }{\left( \omega^{2}_{\ch{Si}} -\omega^{2} -i\omega\gamma_{\ch{Si}} \right)\left( \omega^{2}_{\ch{Au}} -\omega^{2} -i\omega\gamma_{\ch{Au}} \right)-\Omega^{4}}\tilde{f} \label{eq:EquationOfMotionSolutionAu}
\end{align}
\end{subequations}

%% file: Equations1/eq_SurfaceConductivity.tex
\begin{equation}
	\sigma_{e}= -in_{s}\omega\left[  \dfrac{2\Omega^2+\left( \omega^{2}_{\ch{Au}} -\omega^{2} -i\omega\gamma_{\ch{Au}} \right)+\left( \omega^{2}_{\ch{Si}} -\omega^{2} -i\omega\gamma_{\ch{Si}} \right) }{\left( \omega^{2}_{\ch{Si}} -\omega^{2} -i\omega\gamma_{\ch{Si}} \right)\left( \omega^{2}_{\ch{Au}} -\omega^{2} -i\omega\gamma_{\ch{Au}} \right)-\Omega^{4}}\right]
\label{eq:SurfaceConductivity}
\end{equation}

%% file: Equations1/eq_TransmissionCoefficient.tex
\begin{equation}
T=\dfrac{2}{2+Z_{0}\sigma_{e}}
\label{eq:TransmissionCoefficient}
\end{equation}

%% file: Figures1/fig_Transmission_fitting.tex
\begin{figure}[h!]
 \centering
  \includegraphics[width=0.7\textwidth]{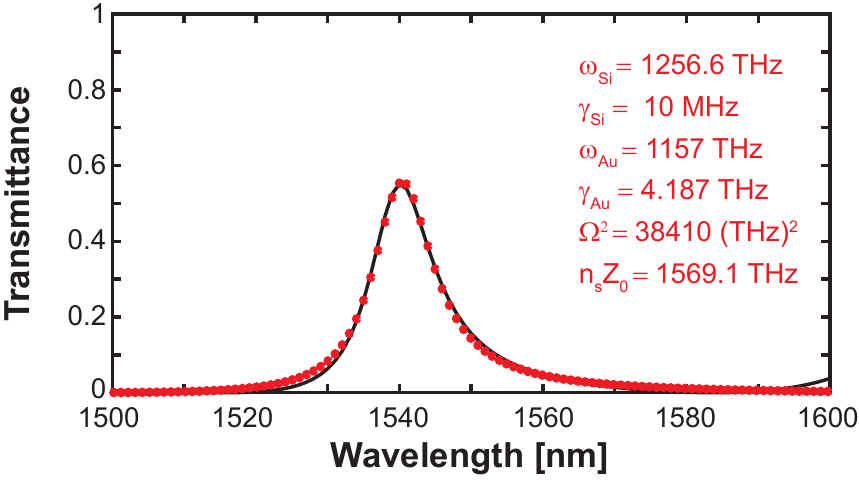}
\caption{Fitting the numerically obtained transmittance (black curve) with radiative coupled-oscillator model (red dots). The fit parameters are shown in the insert.}
\label{fig:Transmission_fitting}
\end{figure}

%% file: Equations1/eq_inlineCurrentDensity.tex
\begin{equation}
\vec{\mathbf{J}}(r) = i\omega(\varepsilon\mts{particle}-\varepsilon\mts{host})\vec{\mathbf{E}}(r)
\label{eq:inlineCurrentDensity}
\end{equation}

%% file: Equations1/eq_ElectricDipole.tex
\begin{equation}
p_j=\frac{i}{\omega}\int\limits_v J_j\;{\mathrm{d}}v
\label{eq:ElectricDipole}
\end{equation}

%% file: Equations1/eq_MagneticDipole.tex
\begin{equation}
m_j=\frac{1}{2}\int\limits_v ({\mathbf{\vec{r}}}\times{\mathbf{\vec{J}}})_j\;{\mathrm{d}}v
\label{eq:MagneticDipole}
\end{equation}

%% file: Equations1/eq_ToroidalEDipole.tex
\begin{equation}
T^{(e)}_{j}=\frac{1}{10}\int\limits_v [({\mathbf{\vec{J}}}\cdot{\mathbf{\vec{r}}})r_{j}-2r^{2}J_{j}]\;{\mathrm{d}}v
\label{eq:ToroidalEDipole1}
\end{equation}

\begin{equation}
T^{(2e)}_{j}=\frac{1}{280}\int\limits_v [3r^4J_j - 2r^2(\mathbf{\vec{r}}\cdot \mathbf{\vec{J}})r_j]\;{\mathrm{d}}v
\label{eq:ToroidalEDipole2}
\end{equation}

%% file: Equations1/eq_ToroidalMDipole.tex
\begin{equation}
T^{(m)}_{j}=\frac{i\omega}{20}\int\limits_v r^2({\mathbf{\vec{r}}}\times{\mathbf{\vec{J}}})_{j}\;{\mathrm{d}}v
\label{eq:ToroidalMDipole}
\end{equation}

%% file: Equations1/eq_ElectricQuadrapole.tex
\begin{equation}
{\bar{\bar{Q}}}^{(e)}_{jk}=\frac{i}{\omega}\int\limits_v [r_jJ_k+r_kJ_j -\frac{2}{3}\delta_{jk}({\mathbf{\vec{r}}}\cdot{\mathbf{\vec{J}}})]\;{\mathrm{d}}v
\label{eq:ElectricQuadrapole}
\end{equation}

%% file: Equations1/eq_MagneticQuadrapole.tex
\begin{equation}
{\bar{\bar{Q}}}^{(m)}_{jk}=\frac{1}{3}\int\limits_v [({\mathbf{\vec{r}}}\times{\mathbf{\vec{J}}})_jr_{k} +({\mathbf{\vec{r}}}\times{\mathbf{\vec{J}}})_{k}r_j]\;{\mathrm{d}}v
\label{eq:MagneticQuadrapole}
\end{equation}

%% file: Equations1/eq_ToroidalEQuadrapole.tex
\begin{equation}
{\bar{\bar{T}}}^{(Qe)}_{jk}=\frac{1}{42}\int\limits_v [4({\mathbf{\vec{r}}}\cdot{\mathbf{\vec{J}}})r_{j}r_{k} +2({\mathbf{\vec{J}}}\cdot{\mathbf{\vec{r}}})r^{2}\delta_{jk}-5r^{2}(r_{j}J_{k}+r_{k}J_{j}) ]\;{\mathrm{d}}v
\label{eq:ToroidalEQuadrapole}
\end{equation}

%% file: Equations1/eq_ToroidalMQuadrapole.tex
\begin{equation}
{\bar{\bar{T}}}^{(Qm)}_{jk}=\frac{i\omega}{42}\int\limits_v r^2[r_j({\mathbf{\vec{r}}}\times{\mathbf{\vec{J}}})_{k} +({\mathbf{\vec{r}}}\times{\mathbf{\vec{J}}})_{j}r_k]\;{\mathrm{d}}v
\label{eq:ToroidalMQuadrapole}
\end{equation}

%% file: Equations1/eq_PScatEDipole.tex
\begin{equation}
P^{(e)}\mts{scat}=\dfrac{k^{4}\sqrt{\varepsilon\mts{host}}}{12\pi \varepsilon^2_0 c\mu_0}\sum\limits_{j=1}^3\vert p\vert_j^2
\label{eq:PScatEDipole}
\end{equation}

%% file: Equations1/eq_PScatTEDipole.tex
\begin{equation}
P^{(Te)}\mts{scat}=\dfrac{k^{4}\sqrt{\varepsilon\mts{host}}}{12\pi \varepsilon^2_0 c\mu_0}\sum\limits_{j=1}^3\lvert \dfrac{ik\varepsilon\mts{host}}{c}T^{(e)}_{j} + \dfrac{ik^{3}\varepsilon\mts{host}^{2}}{c}T^{(2e)}_{j} \rvert ^2
\label{eq:PScatTEDipole}
\end{equation}

%% file: Equations1/eq_PScatMDipole.tex
\begin{equation}
P^{(m)}\mts{scat}=\dfrac{k^{4}\sqrt{\varepsilon^{3}\mts{host}}}{12\pi \varepsilon_0 c}\sum\limits_{j=1}^3\lvert m_{j} + \dfrac{ik\varepsilon\mts{host}}{c}T^{(m)}_{j} \rvert ^2
\label{eq:PScatMDipole}
\end{equation}

%% file: Equations1/eq_PScatEQpole.tex
\begin{equation}
P^{(Qe)}\mts{scat}=\dfrac{k^{6}\sqrt{\varepsilon^{3}\mts{host}}}{160\pi \varepsilon^{2}_0 c\mu_0}\sum\limits_{k=1}^3\sum\limits_{j=1}^3\lvert {\bar{\bar{Q}}}^{(e)}_{jk} + \dfrac{ik\varepsilon\mts{host}}{c}{\bar{\bar{T}}}^{(Qe)}_{jk} \rvert ^2
\label{eq:PScatEQpole}
\end{equation}

%% file: Equations1/eq_PScatMQpole.tex
\begin{equation}
P^{(Qm)}\mts{scat}=\dfrac{k^{6}\sqrt{\varepsilon^{5}\mts{host}}}{160\pi \varepsilon_0 c}\sum\limits_{k=1}^3\sum\limits_{j=1}^3\lvert {\bar{\bar{Q}}}^{(m)}_{jk} + \dfrac{ik\varepsilon\mts{host}}{c}{\bar{\bar{T}}}^{(Qm)}_{jk} \rvert ^2
\label{eq:PScatMQpole}
\end{equation}

%% file: Equations1/eq_scatteringSection.tex
\begin{equation}
\sigma\mts{scat}=2\sqrt{\dfrac{\mu_0}{\varepsilon_0\varepsilon\mts{host}}}\dfrac{P\mts{scat}}{\lvert \mathbf{\vec{E}}\mts{inc}\rvert^2}
\label{eq:scatteringSection}
\end{equation}

%% file: Equations1/eq_ScatteringEfficiency.tex
\begin{equation}
Q\mts{scat}=\dfrac{\sigma\mts{scat}}{\sigma\mts{geom}}
\label{eq:ScatteringEfficiency}
\end{equation}

%% file: Figures1/fig_Filters_multipole_expansion.tex
\begin{figure}[h!]
 \centering
  \includegraphics[width=0.7\textwidth]{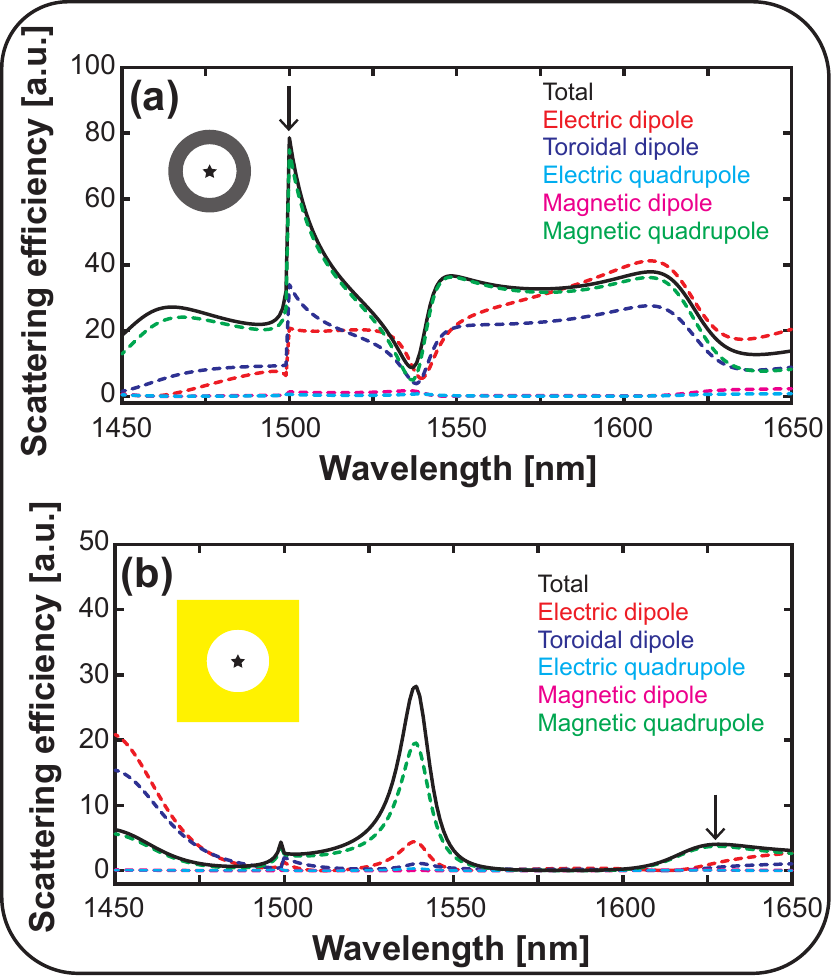}
\caption{Scattering efficiency of the CDMM filter. Modal decomposition of the currents in (a) the a-\ch{Si} ring, (b) the \ch{Au} layer. The black star locates the origin. a-\ch{Si} has a sharp scattering at 1500 nm, \ch{Au} layer has two scattering peaks, one at 1540 nm and one at 1629 nm.}
\label{fig:Filters_multipole_expansion}
\end{figure}

%% file: Figures1/fig_E_Field_1500nm_1629nm.tex
\begin{figure}[h!]
 \centering
  \includegraphics[width=1\textwidth]{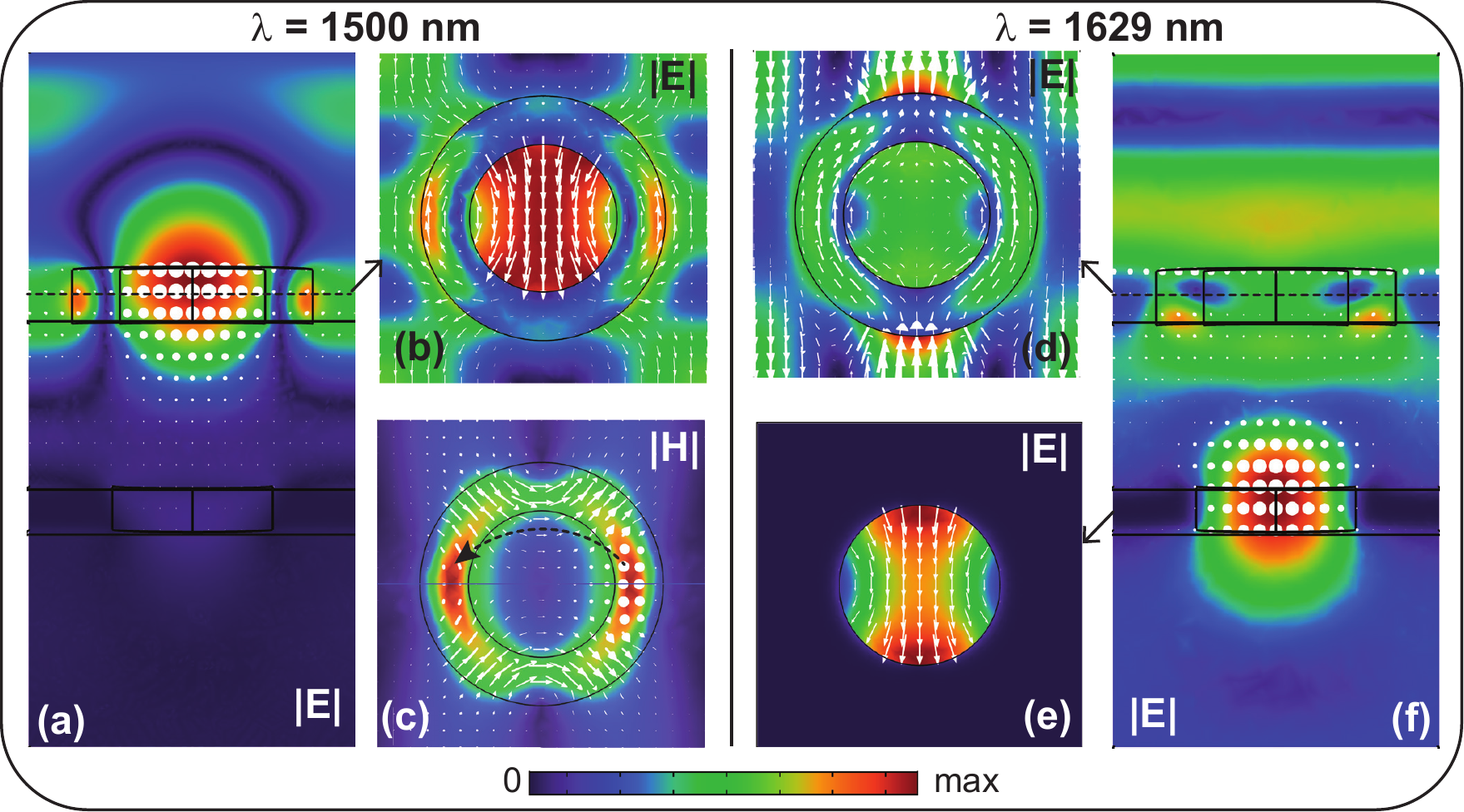}
\caption{Moduli of electric fields of the CDMM filter at 1500 nm along (a) $xz$ plane and (b) $xy$ plane along the center of the a-\ch{Si} ring. (c) The corresponding magnetic field moduli at the $xy$ plane. Moduli of electric fields of the CDMM filter at 1629 nm along (d) $xy$ plane along the center of the a-\ch{Si} ring, (e) $xy$ plane along the center of the \ch{Au} layer, and (f) $xz$ plane. The white arrows, with lengths proportional to the magnitude, indicate the field components. The normally incident planewave is polarized along $y$-axis.}
\label{fig:E_Field_1500nm_1629nm}
\end{figure}

%% file: Figures1/fig_Transmission_offset.tex
\begin{figure}[h!]
 \centering
  \includegraphics[width=1\textwidth]{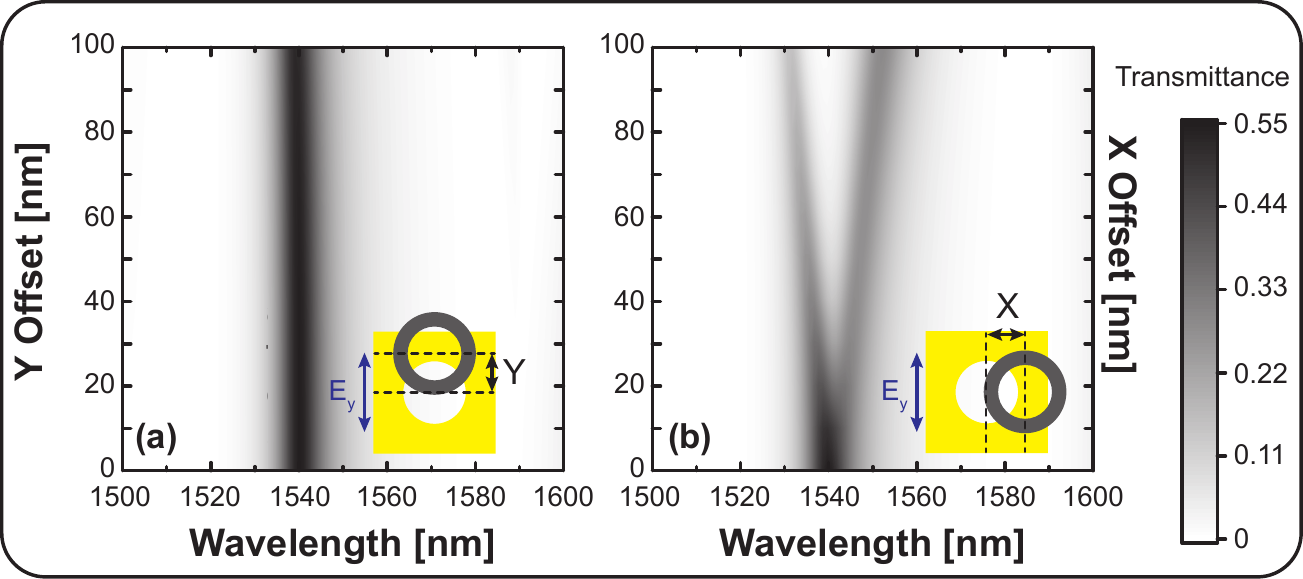}
\caption{Simulated Effects of (a) $Y$ offset and (b) $X$ offset between the a-\ch{Si} ring and the \ch{Au} layer of the CDMM filter on the transmittance. The normal incident planewave illumination is polarized along $y$ axis as shown in the insets.}
\label{fig:Transmission_offset}
\end{figure}

%% file: Figures1/fig_Transmission_Robustness.tex
\begin{figure}[h!]
 \centering
  \includegraphics[width=1\textwidth]{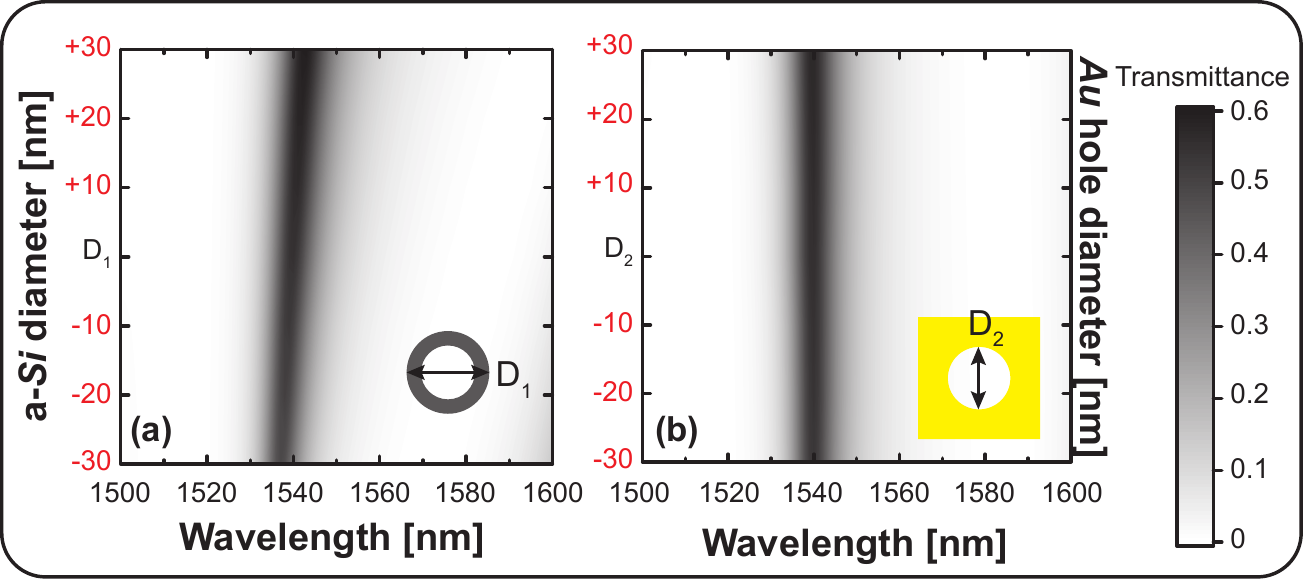}
\caption{Robustness of the CDMM filter. Simulated effects of changes in the diameter of (a) a-\ch{Si} ring and (b) subwavelength holes in \ch{Au} layer of the CDMM filter on the transmittance. The normal incident planewave illumination is polarized along $y$ axis. The ideal diameters are represented by $D_1$ and $D_2$ respectively, while the values in red denote positive/negative changes in nm in the diameter.}
\label{fig:Transmission_Robustness}
\end{figure}

%% file: Figures1/fig_Angle_sweep.tex
\begin{figure}[h!]
 \centering
  \includegraphics[width=0.7\textwidth]{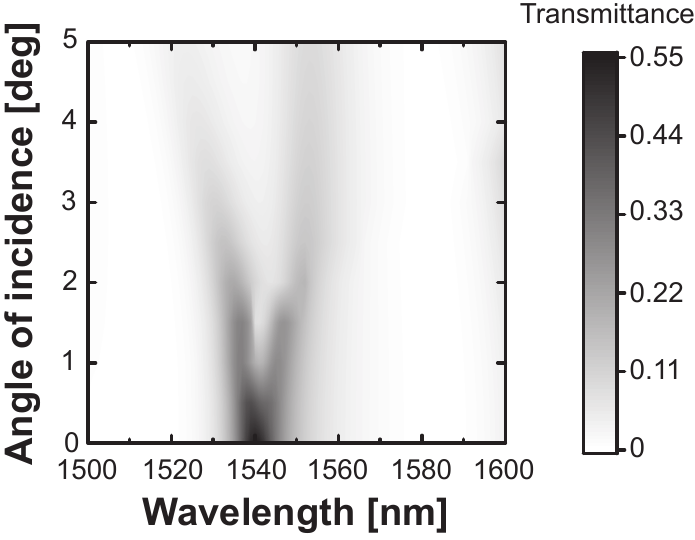}
\caption{Simulated dependence of the CDMM filter transmittance on the angle of incidence for a TE polarized planewave illumination. }
\label{fig:Angle_sweep}
\end{figure}

%% file: Figures1/fig_Period_Sweep.tex
\begin{figure}[h!]
 \centering
  \includegraphics[width=0.7\textwidth]{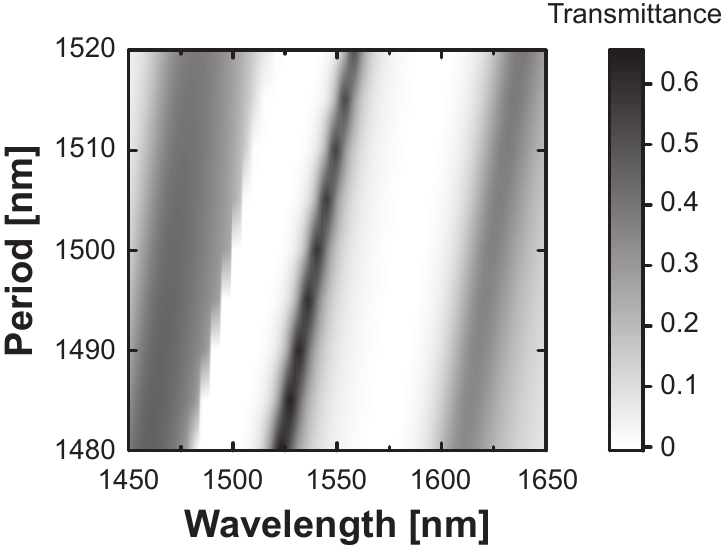}
\caption{Simulated dependence of the peak transmittance on the periodicity ($\Lambda$) of the unit cell for a normal incidence y-polarized illumination. All other design parameters are fixed.}
\label{fig:Period_Sweep}
\end{figure}